\documentclass{iopart}
\usepackage{iopams}
\usepackage{cite}
\usepackage{epsfig}
\usepackage{times}
\usepackage{graphics}
\usepackage{graphicx}
\usepackage{bm}
\newcommand{\UQ}{ARC Centre of Excellence for Quantum-Atom Optics, School of Physical Sciences, University of Queensland, Brisbane QLD 4072, Australia.}

\newcommand{\UCA}{Ultra-Cold Atoms Group, Department of Physics, University of Otago, Dunedin, New Zealand.}
\newcommand{\TAD}{New Zealand Foundation for Research, Science and Technology contract TAD-1054}
\newcommand{\VSO}{Victoria University Scholarships Office}
\newcommand{\CWGFund}{Marsden Fund of the Royal Society of New Zealand contract PVT-202}
\newcommand{\ARC}{Australian Research Council}

\newcommand{\DP}[2]{\frac{\bar{\delta}#1}{\bar{\delta}#2}}
\newcommand{\DDP}[1]{\frac{\bar{\delta}}{\bar{\delta}#1}}
\newcommand{\DPDP}[2]{\frac{\bar{\delta}^2}{\bar{\delta}#1\bar{\delta}#2}}

\newcommand{\bra}[1]{\langle#1|}

\newcommand{\EQ}[1]{\begin{eqnarray}#1\end{eqnarray}}
\newcommand{\mbf}[1]{\mathbf{#1}}
\newcommand{\av}[1]{\langle#1\rangle}
\newcommand{\myint}[1]{\int d^3 #1\;}

\newcommand{\QQ}{{\cal Q}}
\newcommand{\PP}{{\cal P}}
\newcommand {\trNC}[1]{{\rm tr}_{NC}\left(#1\right)}
\newcommand {\ital}[1] {{\em #1}}
\begin{document}
\title{The stochastic Gross-Pitaevskii equation: III}
\author{A.~S. Bradley$^1$ and C.~W. Gardiner$^2$}
\address{$^1$\UQ}
\address{$^2$\UCA}
\date{\today}
\begin{abstract}
A generalised stochastic Gross-Pitaevskii equation describing a partially condensed trapped Bose gas with rotating thermal component is presented. We elucidate the manner in which the rotation changes the role of the high energy cutoff and introduces centrifugal effects in the classical field evolution.
The rotation of the cloud means that Bose-enhanced collision processes occur preferentially into states with the same angular momentum as the thermal cloud, thus favouring vortex formation. We use the formalism to obtain a first principles theory of vortex lattice formation caused by thermodynamic instability.
\end{abstract}
\maketitle
\section{Introduction}
\subsection{Background}
The experimental achievement of dilute gaseous Bose-Einstein condensation~\cite{Anderson1995} began a new era of intense interest in atomic, molecular and optical physics. Among their many fascinating properties, trapped dilute gas Bose-Einstein condensates (BECs) exhibit the properties of superfluidity in a precisely controllable and theoretically quantifiable manner~\cite{Dalfovo1999}. At very low temperatures the trapped gas behaves like an ideal superfluid described by a macroscopic order parameter. 
Two persistent themes throughout the field are the Gross-Pitaevskii (GP) description of the condensate~\cite{Pitaevskii1961,Gross1961,Gross1963}, and the Bogoliubov theory of its quasiparticle excitations~\cite{Bogoliubov1947}. The GP theory is an approximate zero temperature description of atomic BEC that accurately predicts a wide range of BEC behaviour, including ground states, dynamics, vortices and solitons~\cite{Dalfovo1999}. However, it neglects some important quantum mechanical effects which we discuss below. The first quantum correction to the classical field theory of Gross and Pitaevskii\footnote[1]{The use of the word `classical' here requires some justification; a full discussion is given in \sref{truncatedWig}} is provided by the Bogoliubov theory which gives a quantum description of weak excitations near zero temperature -- the {\em quasiparticle excitations} of the condensate.
\subsection{Finite temperature effects and vortex formation mechanisms}
However, there are interesting dynamics that require a \ital{high temperature} theory for their description, which necessarily must go beyond the GP and Bogoliubov theories. By high temperature we mean that the temperature is significant compared to the single particle energies of the trapping potential. For a harmonic potential with angular frequency $\omega$ the condition is
\EQ{
\hbar\omega \ll k_BT,
}
where $k_B$ is Boltzmann's constant, implying that thermal fluctuations will play an important role in the system. 
The process of vortex formation in BEC involves aspects of thermal, nonlinear and many body quantum physics, and our aim is to provide a theory that correctly represents the important physics of each of these.
\subsubsection{Stirring a condensate}
Several experimental groups have generated vortices in a trapped BEC using a rotating elliptical perturbing potential effected with a focused detuned laser beam~\cite{Chevy2000,Onofrio2000,Madison2000,Hodby2001,Madison2001,Abo-Shaeer2001,Abo-Shaeer2002}. The physical mechanism is the introduction of a symmetry breaking perturber which couples the ground and excited states of the condensate. Perturbative analyses of stirring~\cite{Dalfovo2000,Fetter2001b,Anglin2001,Muryshev2001} have found the critical rotational speed at which dynamical instability sets in leading to growth of the vortex state, and shown that the critical frequency is given by the angular form of the Landau critical velocity for superfluid dissipation. Thus it would appear that dissipation is essential to the process, as it is required for the system to surmount the energy barrier which otherwise prevents vortex penetration from the condensate edge~\cite{Fetter2001}.
\par
In a non-perturbative simulation of a localized stirrer based on the Gross-Pitaevskii equation~\cite{CarradocDavies1999}, it was found that a kind of nonlinear Rabi cycling occurs where the vortex periodically cycles from infinity into the condensate interior. In the absence of dissipation a stable lattice does not form.
The physical role of the stirrer in experiments, therefore, must be both to excite multi-pole modes of the condensate, and to create a rotating thermal component which can provide the requisite dissipation.
\par
The physical mechanism arises from the fact that a condensate containing a vortex is thermodynamically unstable in the laboratory frame, so that the system energy can be lowered by irreversible excitation of the anomalous mode~\cite{Isoshima2003}. In the presence of dissipation the vortex spirals out of the condensate. However, for this to occur the dissipation must arise in the lab frame, corresponding physically to a condensate immersed in a stationary thermal cloud. Conversely, if the cloud is rotating faster than the critical frequency for vortex nucleation, the lowest energy state in the co-rotating frame is a condensate containing a stable vortex lattice~\cite{Fetter2001}.
\subsubsection{Cooling a rotating cloud}
It has long been appreciated that phase defects can be trapped during a phase transition~\cite{Kibble1976,Zurek1985,Chuang1991}, thus providing another vortex formation mechanism. This has been demonstrated theoretically for Bose-Einstein condensation by Anglin and Zurek~\cite{Anglin1999}, Marshall~\etal~\cite{Marshall1999} and Davis~\etal~\cite{Davis2002}, but the mechanism has not yet been observed in BEC experiments.
\par
In a series of elegant experiments, Haljan~\etal~\cite{Haljan2001} used a novel technique to create large vortex lattices. Unlike the other nucleation experiments at that time which acted on a pre-formed condensate with some kind of mechanical stirrer, this experiment involved evaporatively cooling a rotating thermal cloud to quantum degeneracy. In this way, it became possible to directly nucleate a condensate containing a disordered array of vortices with the same sense of circulation as the thermal cloud; nevertheless, direct evidence that this process involves defect pinning during the phase transition has not been forthcoming.
\par
Thus far there has been no theory of the formation of vortex lattices that includes the effects of thermal and quantum fluctuations, and, perhaps more importantly, is capable of describing the rotating condensation experiments of Haljan~\etal~\cite{Haljan2001}. The development of such a theory has, in large part, been the motivation for this work. 
\subsubsection{Including dissipation}
One theoretical approach to studying formation dynamics has been to introduce some form of phenomenological damping term in the GPE~\cite{Choi1998,Tsubota2002}, which is intended to model the effect of a rotating thermal cloud. In the absence of an exact microscopic derivation several different approaches have been taken as to the form such a damping term should take. Tsubota \etal~\cite{Tsubota2002} introduced a small imaginary coefficient in the time derivative of the GPE, in the rotating frame of the thermal cloud. Gardiner~\etal~\cite{SGPEI} derived a similar equation, but from a more physical point of view, by replacing the condensate chemical potential that arises in the GPE from a local energy conserving kinetic theory with a time derivative, leading to the Gardiner-Anglin-Fudge equation. Both of these approaches lead to the same physics as the form of the vortex growth equation we derive in \sref{sec:svge} of this paper\footnote[1]{The exact equivalence arises in the limit of low damping.}, which is found from a clear microscopic derivation within controlled approximations. 
\subsubsection{Hamiltonian chaos, Kolmogorov turbulence and the Wigner representation}
A paradox arises when considering the mechanism of vortex formation via mechanical stirring. On one hand, it seems clear that some kind of dissipation is required which would be provided by a rotating thermal component. On the other hand, time evolution in quantum mechanics is unitary, so that if we were able to fully simulate the quantum many body problem for stirring a BEC initially in a ground state of the trap in the laboratory frame, we would expect to find that the system {\em never} reaches equilibrium in the rotating frame of the stirrer.
\par
Putting aside the origin of dissipation in quantum mechanics which can often be trivially resolved by recourse to the law of large numbers, a similar paradox arises in the {\em Gross-Pitaevskii} theory, which is also a Hamiltonian theory. In practice one observes that the dynamics are irreversible -- non-equilibrium initial states will evolve into quasi-stationary states with ergodic properties, wherein time evolution generates a thermal ensemble~\cite{Blakie2005a}. An important feature of the GPE is that the motion can exhibit features of classical Hamiltonian chaos~\cite{Thommen2003}. It is therefore possible that a kind of \ital{internal} dissipation can provide a mechanism for vortex nucleation under the right circumstances. This possibility has been considered in the work of Lobo~\etal~\cite{Lobo2003} who used the classical field method to study the effect of a rotating drive on a BEC. The authors found that in the absence of any external dissipation or initial noise the nonlinear turbulence of the classical field evolution in the vicinity of a dynamical instability~\cite{Sinha2001} was sufficient to relax the classical field toward equilibrium -- although the relaxation occurs over a rather long timescale.
\par
The resolution of this paradox comes from the fact that the classical field evolution of the GPE does not provide a complete physical description of the many body quantum system of interest. In fact, we can expect the Kolmogorov spectrum of turbulence exhibited by the GPE near the instability~\cite{Parker2005} to be associated with the generation of a physical component of the gas that {\em cannot} be described within GP theory -- a thermal component consisting of many weakly occupied modes. In Kolmogorov turbulence energy cascades from long to short wavelengths, and it is the turbulent dynamics at short wavelengths which are difficult to observe, difficult to simulate accurately within GP theory, and the reason why the GP evolution leads to dissipation.
\par 
Indeed, the method of dynamical evolution of the GPE with initial noise arises from a Wigner representation of the quantum field, provided that terms beyond the GPE in the evolution which account for important quantum mechanical effects are {\em truncated}~\cite{Steel1998}. The truncation is usually justified if all of the modes included in the GP description are {\em highly occupied}\footnote[1]{Recent work by Norrie \etal~\cite{Norrie2006} shows that it is possible to relax the high mode occupation criteria somewhat, in particular, for the condensate collision scenario of~\cite{Norrie2005} where many vacuum modes are included in the description.}, which is clearly not the case during a dynamical instability since many high energy modes are excited. Thus, the {\em physical} resolution of the paradox must come about from a theory that accurately describes the thermal cloud, and in particular its formation and influence on the coherent component of the gas.
\par 
There is a distinct need, then, for a theory that accurately describes the rotating thermal component of a trapped Bose gas, and which includes the stimulated collisions between the condensate and noncondensate atoms. In this work we do not address the {\em formation} of the thermal cloud, but we instead go some way in the direction of a theory of vortex formation by providing a theory of the {\em influence} of the thermal cloud, a description which is necessarily {\em stochastic} in its formulation. 
\subsection{SGPE theory}
One particularly successful way of handling both finite temperature and many body effects consists of treating the trapped Bose gas as an open system. The finite temperature Bose gas treated as a condensate coupled to a thermal reservoir which generates dissipative effects. The first derivation of a generalised stochastic Gross-Pitaevskii equation (SGPE) from this point of view for trapped BEC was provided by Stoof~\cite{Stoof1999,Stoof2001} using functional techniques, starting from a Keldysh path integral formulation of Quantum Field Theory. The differences between the theory of \cite{Stoof1999} and the present SGPE theory are discussed in \cite{SGPEII}, and we refer the reader to this paper for detailed discussion; however, the most notable difference is the absence of a particular kind of interaction between the condensate and thermal cloud, which we refer to as the \ital{scattering term}, which will be described in \sref{sec:bandInt}. This term does not directly change the occupation of the condensate band, and so often its effect can be justifiably neglected. It \ital{does} however couple to superfluid flows in the condensate band, so that, for example, we may expect it to play an important role in vortex motion.
Stoof's approach has certain advantages, most notably, since it is based on a Lagrangian formulation it is amenable to variational techniques, and a Gaussian variational method has been developed to great effect by Duine \etal~\cite{Duine2002,Duine2004}. Nevertheless, as we have noted, there are certain effects which are difficult to capture using path integral methods, and for this reason we use the open systems approach developed in quantum optics~\cite{QO,QN}.
\par 
In this paper we derive a generalised SGPE describing a rotating partially condensed trapped Bose gas, by adapting the theory developed by Gardiner~\etal~\cite{SGPEI,SGPEII}. In order to handle both a coherent condensed fraction {\em and} a high energy incoherent fraction, the SGPE formalism separates the partially condensed system
into a low energy subspace of modes -- the {\em condensate band},
and its high energy counterpart -- the 
{\em noncondensate band}. The latter is treated as being thermalised, thus playing the role of a reservoir which damps the dynamics of the condensate band -- in the rotating frame of the thermal cloud. An important feature of the approach, which also distinguishes it from that of \cite{Stoof1999}, is that the condensate band contains both the condensate {\em and} its excitations. 
\par
Using this approach we obtain stochastic equations of motion similar to those describing the nonrotating scenario of~\cite{SGPEII}. In the final analysis we show that the rotation generates additional centrifugal terms in the description of the condensate band and the thermal cloud. Since there are also some corrections to the derivation and final form supplied in \cite{SGPEII}, our approach in the present paper is to provide a reasonably self contained generalisation of that work and to supply a number of necessary corrections. 
\subsection{Qualitative effect of the rotation}
There are two scenarios of interest in which a non-condensed Bose gas in some state of motion  may come into thermal equilibrium. 
\begin{itemize}
\item In the first, the gas can have a constant center of mass motion in the laboratory, either confined to an atomic waveguide~\cite{JJHThesis}, or in a translating harmonic trap. It is well known that within classical statistics the thermodynamic properties of a uniformly translating gas are unaltered by the motion~\cite{LandLSM1} -- a consequence of Galilean invariance. This invariance trivially carries over to quantum statistics, and the dynamics of Bose-Einstein condensation for a uniformly translating gas are thus completely unaltered by the center of mass motion: the equations of motion are identical to those for the stationary system, modulo a shift in center of mass momentum, after a Galilean transformation into the moving frame. 
\item The second situation of interest occurs when the gas is confined in a cylindrically symmetric trap, and has non-zero angular momentum. The gas is thus at rest in a frame rotating about the symmetry axis of the trap. In this case the motion of the cloud has a profound effect on the dynamics of Bose-Einstein condensation.  
\end{itemize}
A central feature of what follows is that the theory is developed in the frame co-rotating with the thermal cloud. Within classical statistics the only effect of rotation is to alter the dynamics of the thermal cloud such that it evolves in a centrifugally modified effective potential~\cite{LandLSM1}. 
\begin{figure}[!htb]{
\begin{center} 
\includegraphics[width=8cm]{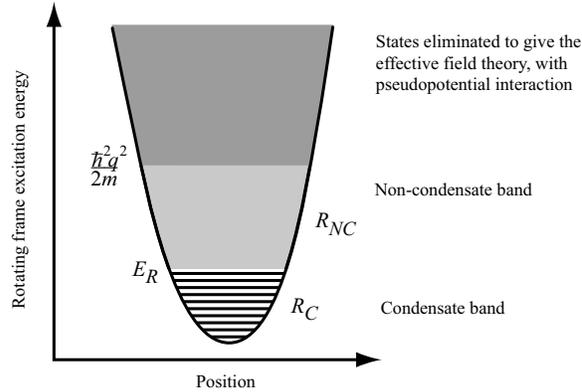}
\caption{Schematic of the separation into condensate and noncondensate bands. States beneath the energy $E_R$, defined in the rotating frame, form the description of the \ital{condensate band} which warrant a quantum mechanical description, and the states above $E_R$ form the \ital{noncondensate band} which consists of high energy, approximately thermal states. The S-wave scattering description of the collisions means that states with energies above $\hbar^2q^2/2m$ in the rotating frame are also eliminated from the theory~\cite{SGPEII}.
\label{band_schematic}}
\end{center}}
\end{figure}
Within quantum statistics, however, there are important centrifugal terms which appear in the growth and scattering terms of the SGPE which we derive. It will be shown that the full theory of~\cite{SGPEII} can be used with the appropriate changes to the single particle Hamiltonian and the thermal distributions in the rotating frame. 
\section{The system}
\subsection{The cold-collision Hamiltonian}
In the cold-collision regime described by S-wave scattering length $a$~\cite{Dalfovo1999} the second-quantised many body Hamiltonian for a dilute Bose gas confined by a trapping potential $V(\mbf{x})$, formulated in the rotating frame defined by angular frequency $\mbf{\Omega}$ is~\cite{Pitaevskii&Stringari}
\EQ{\label{fullHamil}
H=H_{\rm sp}+H_I
}
-
where the {\em single-particle} Hamiltonian is
\EQ{
H_{\rm sp}&=&\myint{\mbf{x}}\psi^\dag(\mbf{x})\left(-\frac{\hbar^2\nabla^2}{2m}+i\hbar\mbf{\Omega}\cdot (\mbf{x}\times\nabla)+V(\mbf{x})\right)\psi(\mbf{x})
}
and the {\em interaction Hamiltonian} is
\EQ{\label{ccint}
H_I=\frac{u}{2}\myint{\mbf{x}}\psi^\dag(\mbf{x})\psi^\dag(\mbf{x})\psi(\mbf{x})\psi(\mbf{x}),
}
and where $u=4\pi\hbar^2a/m$, $m$ is the atomic mass and $a$ is the S-wave scattering length. 
As shown schematically in \fref{band_schematic}, the states of the trapped system are divided into the {\em condensate band} of states with energy beneath the cut-off $E_R$, and the remaining {\em noncondensate band} of high energy states. The separation into distinct bands allows the two regions to be treated using different techniques. The high energy states are to be treated using the master equation techniques of quantum optics, and will physically play the role of a thermal reservoir for the low energy condensate band. The condensate band modes are assumed to be, at least, moderately occupied so that some of the quantum mechanical features of the condensate band may be consistently neglected. The field operator is 
\EQ{\label{divide_field}
\psi(\mathbf{x})=\phi(\mathbf{x})+\psi_{\rm NC}(\mathbf{x})\equiv\PP\psi(\mathbf{x})+\QQ\psi(\mathbf{x}),
}
where the noncondensate
field $\psi_{\rm NC}(\mathbf{x})\equiv\QQ\psi(\mbf{x})$ describes the high energy thermal modes, and the condensate band is described by the field $\phi(\mbf{x})\equiv\PP\psi(\mbf{x})$. The the orthogonal projectors $\PP+\QQ=1$ are defined with respect to the single particle basis of the trap. 
The condensate band projector takes the form
\EQ{\label{Pdef1}
\PP\equiv\sum_n^-|n\rangle\langle n|
}
where the bar denotes the cut-off and $n$ represents all quantum numbers required to index the eigenstates of the single particle Hamiltonian. In the spatial representation the eigenstates are denoted by $Y_n(\mbf{x})\equiv\bra{\mbf{x}}n\rangle$ and the action of the projector on an arbitrary wavefunction $\chi(\mbf{x})$ is
\EQ{\label{Pdef2}
\PP\chi(\mbf{x})=\sum_n^-\;Y_n(\mbf{x})\myint{\mbf{z}}Y_n^*(\mbf{z})\chi(\mbf{z}).
}
\subsection{Separation of the Hamiltonian}
Using \eref{divide_field} the full Hamiltonian~\eref{fullHamil} can be written as
\EQ{
H=H_0+H_{I,C}+H_{NC}
}
where $H_0$ involves only condensate band operators, $H_{NC}$ involves only noncondensate band operators and 
\EQ{
H_{I,C}=H_{I,C}^{(1)}+H_{I,C}^{(2)}+H_{I,C}^{(3)}
}
denote interaction terms involving one, two or three condensate band operators:
\EQ{\fl
H_0&=&\myint{\mbf{x}}\phi^\dag(\mbf{x})\left(-\frac{\hbar^2\nabla^2}{2m}+i\hbar\mbf{\Omega}\cdot (\mbf{x}\times\nabla)+V(\mbf{x})\right)\phi(\mbf{x})\nonumber\\\fl
\label{H0def1}&&+\frac{u}{2}\myint{\mbf{x}}\phi^\dag(\mbf{x})\phi^\dag(\mbf{x})\phi(\mbf{x})\phi(\mbf{x})\\\fl
H_{NC}&=&\myint{\mbf{x}}\psi_{NC}^\dag(\mbf{x})\left(-\frac{\hbar^2\nabla^2}{2m}+i\hbar\mbf{\Omega}\cdot (\mbf{x}\times\nabla)+V(\mbf{x})\right)\psi_{NC}(\mbf{x})\nonumber\\\fl
&&+\frac{u}{2}\myint{\mbf{x}}\psi_{NC}^\dag(\mbf{x})\psi_{NC}^\dag(\mbf{x})\psi_{NC}(\mbf{x})\psi_{NC}(\mbf{x})\\\fl
\label{HI1def}
H_{I,C}^{(1)}&=&u\myint{\mbf{x}}\psi_{NC}^\dag(\mbf{x})\psi_{NC}^\dag(\mbf{x})\psi_{NC}(\mbf{x})\phi(\mbf{x})+{\rm h.c.}\\\fl
H_{I,C}^{(2)}&=&u\myint{\mbf{x}}\psi_{NC}^\dag(\mbf{x})\psi_{NC}(\mbf{x})\phi^\dag(\mbf{x})\phi(\mbf{x})+{\rm h.c.}\nonumber\\\fl
\label{HI2def}&&+\frac{u}{2}\myint{\mbf{x}}\psi_{NC}^\dag(\mbf{x})\psi_{NC}^\dag(\mbf{x})\phi(\mbf{x})\phi(\mbf{x})+{\rm h.c.}\\\fl
\label{HI3def}
H_{I,C}^{(3)}&=&u\myint{\mbf{x}}\phi^\dag(\mbf{x})\phi^\dag(\mbf{x})\phi(\mbf{x})\psi_{NC}(\mbf{x})+{\rm h.c.}
}
We note that there are additional terms arising from the single particle Hamiltonian involving one $\psi_{NC}(\mbf{x})$ and one $\phi(\mbf{x})$ but these play no role in the following theory because the noncondensate band operators have no mean field.
\section{Condensate band master equation}
\subsection{Formal derivation of the master equation}
The master equation for the condensate band density operator
\EQ{
\rho_C\equiv\trNC{\rho}
}
is found from the von Neumann equation for the total density operator
\EQ{
\dot{\rho}&=&-\frac{i}{\hbar}[H_{0}+H_I+H_{NC},\rho]\nonumber\\
&\equiv&({\cal L}_{0}+{\cal L}_{I}+{\cal L}_{NC})\rho
}
by defining the density operator projector
\EQ{
\PP_C\rho\equiv\rho_{NC}\otimes\trNC{\rho}\equiv \rho_{NC}\otimes\rho_C,
}
together with its orthogonal projector $\QQ_C=1-\PP_C$; one then uses standard methods~\cite{QN} 
to find the equation of motion for the Laplace transform of $v(t)\equiv\PP_C\rho(t)$
\EQ{\label{vseom}
s\tilde{v}(s)-v(0)&=&({\cal L}_0+\PP_C{\cal L}_I)\tilde{v}(s)\nonumber\\
&&+\PP_C{\cal L}_I[s-{\cal L}_0-{\cal L}_I-{\cal L}_{NC}]^{-1}\QQ_C{\cal L}_I\tilde{v}(s).
}
At this point two approximations are made. Firstly the evolution is approximated by neglecting the interaction superoperator ${\cal L}_I$ in $[\;]^{-1}$, which amounts to treating the interactions as a perturbation to the bare system evolution. After using the Laplace convolution theorem to invert~\eref{vseom}, a Markov approximation is made for the time integral by neglecting the time dependence of $v(t)$ over the support of the integral and extending the upper limit of integration to infinity. The approximation is that the reservoir correlation time is small compared to the timescale of system evolution. The result is
\EQ{\label{veomdef}
\dot{v}(t)&=&({\cal L}_0+\PP_C{\cal L}_I)v(t)\nonumber\\
&&+\left\{\PP_C{\cal L}_I\int_0^\infty d\tau \exp{\left\{({\cal L}_0+{\cal L}_{NC})\tau\right\}}\QQ_C{\cal L}_I\right\}v(t).
}
We now evaluate the terms in \eref{veomdef}.
\subsection{Hamiltonian terms}
The term $({\cal L}_0+\PP_C{\cal L}_I)$ in \eref{veomdef} can be written in terms of an effective {\em condensate band Hamiltonian}
\EQ{\label{HCdefsgpe}
H_C\equiv H_0+H_{\rm forward}
}
where
\EQ{
H_{\rm forward}\equiv 2u\myint{\mbf{x}}\bar{n}_{NC}(\mbf{x})\phi^\dag(\mbf{x})\phi(\mbf{x})
}
and the noncondensate density
\EQ{\label{NCdensity}
\bar{n}_{\scriptstyle NC}(\mbf{x})\equiv\langle\psi^\dag(\mbf{x})\psi(\mbf{x})\rangle\equiv\trNC{\rho_{NC}\psi^\dag(\mbf{x})\psi(\mbf{x})}
}
plays the role of an effective potential. The Hamiltonian term is 
\EQ{\label{Hamrho}
\dot{\rho}_C\big{|}_{\rm Ham}=-\frac{i}{\hbar}[H_C,\rho_C].
}
\begin{figure}[!htb]{
\begin{center} 
\includegraphics[width=8cm]{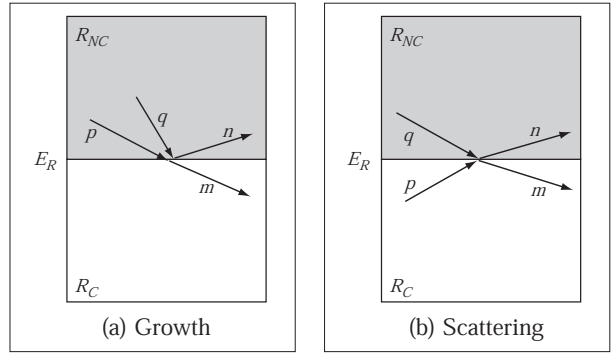}
\caption{Schematic of the processes arising from the interactions between the condensate and noncondensate bands. In (a) two noncondensate band atoms collide. The collision energy is transferred to one of the atoms with the other passing into the condensate band. In (b) a condensate band atom collides with a noncondensate band atom with no change in condensate band population. The time reverse of the two processes also occurs. Figure adapted from~\cite{QKVII}.
\label{process_schematic}}
\end{center}}
\end{figure}
\subsection{Interactions between condensate and noncondensate bands}\label{sec:bandInt}
The master equation terms arising from evaluating~\eref{veomdef} for $H_{I,C}^{(1)}$ and $H_{I,C}^{(2)}$ are
\EQ{\label{Hint1terms}\fl
\dot{\rho}_C\big{|}_{H_{I,C}^{(1)}}&=&\frac{u^2}{\hbar^2}\int d^3\mbf{x}\int d^3\mbf{x^\prime}\int_{-\infty}^0 d\tau\;\Big\{\nonumber\\\fl
\label{corr1}&&\av{\psi_{NC}^\dag(\mbf{x^\prime})\psi_{NC}(\mbf{x^\prime})^2\psi_{NC}^\dag(\mbf{x},\tau)^2\psi_{NC}(\mbf{x},\tau)}
[\phi(\mbf{x},\tau)\rho_C,\phi^\dag(\mbf{x^\prime})]
\nonumber\\\fl
&&+\av{\psi_{NC}^\dag(\mbf{x},\tau)^2\psi_{NC}(\mbf{x},\tau)\psi_{NC}^\dag(\mbf{x^\prime})\psi_{NC}(\mbf{x^\prime})^2}
[\phi^\dag(\mbf{x^\prime}),\rho_C\phi(\mbf{x},\tau)]
\nonumber\\\fl
&&+\av{\psi_{NC}^\dag(\mbf{x^\prime})^2\psi_{NC}(\mbf{x^\prime})\psi_{NC}^\dag(\mbf{x},\tau)\psi_{NC}(\mbf{x},\tau)^2}
[\phi^\dag(\mbf{x},\tau)\rho_C,\phi(\mbf{x^\prime})]
\nonumber\\\fl
&&+\av{\psi_{NC}^\dag(\mbf{x},\tau)\psi_{NC}(\mbf{x},\tau)^2\psi_{NC}^\dag(\mbf{x^\prime})^2\psi_{NC}(\mbf{x^\prime})}
[\phi(\mbf{x^\prime}),\rho_C\phi^\dag(\mbf{x},\tau)]\Big\}
}
\EQ{\label{Hint2terms}\fl
\dot{\rho}_C\big{|}_{H_{I,C}^{(2)}}&=&\frac{u^2}{4\hbar^2}\int d^3\mbf{x}\int d^3\mbf{x^\prime}\int_{-\infty}^0 d\tau\;\Big\{\nonumber\;\;\;\;\;\;\;\;\;\;\;\;\;\\\fl
&&\;\;\;\;\av{\psi_{NC}^\dag(\mbf{x},\tau)\psi_{NC}(\mbf{x},\tau)\psi_{NC}^\dag(\mbf{x^\prime})\psi_{NC}(\mbf{x^\prime})}\nonumber\\\fl &&\;\;\;\;\;\;\;\times[\rho_C\phi^\dag(\mbf{x},\tau)\phi(\mbf{x},\tau),\phi^\dag(\mbf{x^\prime})\phi(\mbf{x^\prime})]\nonumber\;\;\;\;\;\;\;\;\;\;\;\;\;\\\fl
&&+\av{\psi_{NC}^\dag(\mbf{x^\prime})\psi_{NC}(\mbf{x^\prime})\psi_{NC}^\dag(\mbf{x},\tau)\psi_{NC}(\mbf{x},\tau)}\nonumber\\\fl
&&\;\;\;\;\;\;\;\times[\phi^\dag(\mbf{x^\prime})\phi(\mbf{x^\prime}),\phi^\dag(\mbf{x},\tau)\phi(\mbf{x},\tau)\rho_C]\Big\}\;\;\;\;\;\;\;\;\;\;\;\;\;
}
where the time evolved field operators are defined as
\EQ{
\phi(\mbf{x},t)&=&e^{iH_Ct/\hbar}\phi(\mbf{x})e^{-iH_Ct/\hbar}\\
\psi_{NC}(\mbf{x},t)&=&e^{iH_{NC}t/\hbar}\psi_{NC}(\mbf{x})e^{-iH_{NC}t/\hbar}.
}
There are also terms that arise in \eref{veomdef} from a mixture of terms from $H_{I,C}^{(1)}$ and  $H_{I,C}^{(3)}$, and from $H_{I,C}^{(3)}$ alone. All of these contributions can be seen to be prohibited by energy and momentum conservation, which is to say they are \ital{non-resonant}, and as such are very small.
\par
The physical processes corresponding to the $H_{I,C}^{(1)}$ (Growth) and $H_{I,C}^{(2)}$ (Scattering) terms are shown schematically in~\fref{process_schematic}. The growth terms correspond to scattering between condensate and noncondensate atoms whereby particle transfer can take place. The so-called scattering terms involve collisions between condensate and noncondensate atoms which nevertheless conserve the populations in each band. 
\subsection{Reservoir correlation functions}
We now introduce the approximations for the correlation functions of the noncondensate band operators.
\subsubsection{Hartree-Fock factorisation}
For a thermal gas one may use Hartree-Fock factorization to expand high order moments in terms of second order moments. The correlation function \eref{corr1}, for example, becomes
\EQ{\fl\label{gaussf}
\av{\psi_{NC}^\dag(\mbf{x^\prime})\psi_{NC}(\mbf{x^\prime})^2\psi_{NC}^\dag(\mbf{x},\tau)^2\psi_{NC}(\mbf{x},\tau)}
&=&2\av{\psi_{NC}^\dag(\mbf{x^\prime})\psi_{NC}(\mbf{x},\tau)}\nonumber\\\fl
&&\times
\av{\psi_{NC}(\mbf{x^\prime})\psi_{NC}^\dag(\mbf{x},\tau)}\nonumber\\\fl
&&\times
\av{\psi_{NC}(\mbf{x^\prime})\psi_{NC}^\dag(\mbf{x},\tau)}
}
where terms of the form $\av{\psi_{NC}^\dag(\mbf{x^\prime})\psi_{NC}(\mbf{x^\prime})}$ are neglected because they do not lead to energy conserving processes in the master equation.
\subsubsection{Semi-classical description of the thermal cloud in the rotating frame}
The single particle Wigner function for the noncondensate field is defined as
\EQ{
F(\mbf{x},\mbf{K})\equiv\int d^3\mbf{v}\;\av{\psi_{NC}^\dag\left(\mbf{x}+\mbf{v}/2\right)\psi_{NC}\left(\mbf{x}-\mbf{v}/2\right)}e^{i\mbf{K}\cdot\mbf{v}}.
}
In the semiclassical description, the noncondensate band Wigner function for a Bose gas confined in a cylindrically symmetric trap in equilibrium in the rotating frame defined by
$\mbf{\Omega}\equiv \Omega\hat{z}$, with chemical potential $\mu$, takes the form~\cite{LandLSM1,APThesis}
\EQ{\label{Fukdef}
F(\mbf{x},\mbf{K})=\frac{1}{\exp{[(\hbar\omega(\mbf{x},\mbf{K})-\mu)/k_BT]}-1}
}
where 
\EQ{
\hbar\omega(\mbf{x},\mbf{K})&=&\frac{\hbar^2\mbf{K}^2}{2m}-\hbar\mbf{\Omega}\cdot(\mbf{x}\times\mbf{K})+V(\mbf{x})\nonumber\\
\label{Krdef}&=&\frac{\hbar^2}{2m}(\mbf{K}-m\mbf{\Omega}\times\mbf{x}/\hbar)^2+V_{\rm eff}(\mbf{x}),
}
and we have introduced the \ital{effective potential}
\EQ{\label{Veffgrowth}
V_{\rm eff}(\mbf{x})&\equiv&V(\mbf{x})-\frac{m(\mbf{\Omega}\times\mbf{x})^2}{2}\\
&=&\frac{m(\omega_r^2-\Omega^2)^2r_\perp^2}{2}+\frac{m\omega_z^2z^2}{2},
}
where $r_\perp=\sqrt{x^2+y^2}$ is the distance from the symmetry axis of the trap. From~\eref{Krdef} we see that, in terms of the rotating frame momentum, the rotating frame semi-classical distribution is a lab frame equilibrium distribution with a centrifugally modified effective potential. The Coriolis force thus has no effect on the equilibrium state of the rotating cloud within the semi-classical picture~\cite{LandLSM1}. 
\par 
Introducing the variables $\mbf{u}\equiv(\mbf{x}+\mbf{x}^\prime)/2$, 
$\mbf{v}\equiv\mbf{x}^\prime-\mbf{x}$, we can write
\EQ{\label{rncexp1}\fl
\av{\psi_{NC}^\dag(\mbf{x^\prime})\psi_{NC}(\mbf{x},\tau)}&=&\av{\psi_{NC}^\dag(\mbf{u}+\mbf{v}/2)\psi_{NC}(\mbf{u}-\mbf{v}/2,\tau)}\nonumber\\\fl
&\approx&\frac{1}{(2\pi)^3}\int_{R_{NC}} d^3\mbf{K}\;F(\mbf{u},\mbf{K})e^{-i\mbf{K}\cdot\mbf{v}-i\omega(\mbf{u},\mbf{K})\tau}\\\fl\label{rncexp2}
\av{\psi_{NC}(\mbf{x^\prime})\psi_{NC}^\dag(\mbf{x},\tau)}
&\approx&\frac{1}{(2\pi)^3}\int_{R_{NC}} d^3\mbf{K}\;[1+F(\mbf{u},\mbf{K})]e^{i\mbf{K}\cdot\mbf{v}+i\omega(\mbf{u},\mbf{K})\tau}
}
where we have made a semi-classical approximation for the time development of the noncondensate field operator and used the fact that in the semiclassical limit $F(\mbf{u},\mbf{K})$ is only significant in the noncondensate region, which is given by the condition 
\EQ{\label{Rncw}
\hbar\omega(\mbf{u},\mbf{K})>E_R.
}
We note that the rotating frame cutoff condition is more complicated than the laboratory frame cutoff since the former is defined by a position dependent rotating frame momentum. However, the only dependence of $F(\mbf{u},\mbf{K})$ on $\mbf{K}-m\mbf{\Omega}\times\mbf{x}/\hbar$ is implicit, through $\hbar\omega(\mbf{u},\mbf{K})$. In practice one can introduce an appropriate change to energy variables in the necessary integrals -- effectively eliminating the position dependence, so that there is no difficulty in handling this complication.
\par
Equations \eref{Krdef}, \eref{rncexp1}, and \eref{rncexp2} form our description of the thermal cloud; the lab frame appearance of~\eref{Krdef} means that much of the Quantum Kinetic Theory~\cite{QKI,QKIII,QKV} that has been developed to describe condensate growth also applies in the rotating frame with straightforward modifications. However, due to the rotation there are some important changes to the condensate band description.
\subsection{Master equation}
\subsubsection{Growth terms}
The terms in the master equation that lead to condensate growth are from the interaction Hamiltonian term $H_{I,C}^{(1)}$, corresponding to the master equation term \eref{Hint1terms}. This term will be rewritten using the approximate description of the reservoir correlation functions introduced above; further, we will neglect the principal value part of the time integrals in \eref{Hint1terms} and \eref{Hint2terms}
\EQ{
\int_{-\infty}^{0}d\tau\; \exp{(-i\omega\tau)}=\pi\delta(\omega)+i{\rm P}/\omega\approx\pi\delta(\omega),
}
which is physically consistent with the Markov approximation since it ensures conservation of energy during collisions between condensate and noncondensate atoms.
\par
The growth terms can be written in terms of the amplitudes\footnote[1]{This corrects 
an extra minus sign in the defining equation (56) of \cite{SGPEII}.} 
\EQ{\fl
\label{Gplus}G^{\scriptscriptstyle(+)}(\mbf{u},\mbf{v},\epsilon)&=&\frac{u^2}{(2\pi)^8\hbar^2}\int d^3\mbf{K}_1\int d^3\mbf{K}_2\int d^3\mbf{K}_3\;\delta(\omega_1+\omega_2-\omega_3-\epsilon/\hbar)\nonumber\\\fl
&&\times F(\mbf{u},\mbf{K}_1)F(\mbf{u},\mbf{K}_2)[1+F(\mbf{u},\mbf{K}_3)]e^{-i(\mbf{K}_1+\mbf{K}_2-\mbf{K}_3)\cdot\mbf{v}},
\\\fl
\nonumber\\\fl
\label{Gminus}G^{\scriptscriptstyle(-)}(\mbf{u},\mbf{v},\epsilon)&=&\frac{u^2}{(2\pi)^8\hbar^2}\int d^3\mbf{K}_1\int d^3\mbf{K}_2\int d^3\mbf{K}_3\;\delta(\omega_1+\omega_2-\omega_3-\epsilon/\hbar)\nonumber\\\fl
&&\times[1+F(\mbf{u},\mbf{K}_1)][1+F(\mbf{u},\mbf{K}_2)]F(\mbf{u},\mbf{K}_3)e^{-i(\mbf{K}_1+\mbf{K}_2-\mbf{K}_3)\cdot\mbf{v}},
}
and the condensate band operator
\EQ{\label{Lcfielddef}
L_{C}\phi(\mbf{x})&\equiv&[\phi(\mbf{x}),H_{C}]
}
in the form\footnote[1]{This corrects a misprint in equation (59) of \cite{SGPEII} where $L_C$ appeared in place
of $-L_C$ in the second and third lines of \eref{rhogrowth}. Consequently the identities \eref{ident1} -- \eref{ident3} required to establish the validity of the equilibrium solution appeared incorrectly in \cite{SGPEII}.}
\EQ{\fl
\label{rhogrowth}\dot{\rho}_C\big{|}_{\rm growth}&=&\;\;\;\;\int d^3\mbf{u}\int d^3\mbf{v}\left[\left\{G^{\scriptscriptstyle(-)}(\mbf{u},\mbf{v},L_{ C})\phi(\mbf{u}-\mbf{v}/2)\right\}\rho_{C},\phi^\dag(\mbf{u}+\mbf{v}/2)\right]\nonumber\\\fl
&&-\int d^3\mbf{u}\int d^3\mbf{v}\left[\rho_{ C}\left\{G^{\scriptscriptstyle(-)}(\mbf{u},\mbf{v},-L_{ C})\phi^\dag(\mbf{u}-\mbf{v}/2)\right\},\phi(\mbf{u}+\mbf{v}/2)\right]\nonumber\\\fl
&&+\int d^3\mbf{u}\int d^3\mbf{v}\left[\left\{G^{\scriptscriptstyle(+)}(\mbf{u},\mbf{v},-L_{ C})\phi^\dag(\mbf{u}-\mbf{v}/2)\right\}\rho_{C},\phi(\mbf{u}+\mbf{v}/2)\right]\nonumber\\\fl
&&-\int d^3\mbf{u}\int d^3\mbf{v}\left[\rho_{ C}\left\{G^{\scriptscriptstyle(+)}(\mbf{u},\mbf{v},L_{ C})\phi(\mbf{u}-\mbf{v}/2)\right\},\phi^\dag(\mbf{u}+\mbf{v}/2)\right].
}
\subsubsection{Scattering terms}
In terms of the scattering amplitude
\EQ{
M(\mbf{u},\mbf{v},\epsilon)&=&\frac{2u^2}{(2\pi)^5\hbar^2}\int d^3\mbf{K}_1\int d^3\mbf{K}_2\;\delta(\omega_1-\omega_2-\epsilon/\hbar)\nonumber\\
&&\times F(\mbf{u},\mbf{K}_1)[1+F(\mbf{u},\mbf{K}_2)]e^{i(\mbf{K}_1-\mbf{K}_2)\cdot\mbf{v}}
}
and the operator
\EQ{
U(\mbf{x})=\phi^\dag(\mbf{x})\phi(\mbf{x}),
}
the scattering terms, given by \eref{Hint2terms}, take the form\footnote[2]{Note the different operator ordering here corrects the form in (67) of ~\cite{SGPEII}. The resulting stochastic description is unchanged.}
\EQ{\fl
\dot{\rho}_C\big{|}_{\rm scatt}&=&\int d^3\mbf{u}\int d^3\mbf{v}\;[U(\mbf{u}+\mbf{v}/2),\rho_C\left\{M(\mbf{u},\mbf{v},L_C)U(\mbf{u}-\mbf{v}/2)\right\}]\nonumber\\\fl
&&+\int d^3\mbf{u}\int d^3\mbf{v}\;[ \left\{M(\mbf{u},\mbf{v},-L_C)U(\mbf{u}-\mbf{v}/2)\right\}\rho_C,U(\mbf{u}+\mbf{v}/2)]
}
\subsubsection{Forward-backward relations}
When the noncondensate band is described by the thermal equilibrium form given by \eref{Fukdef}, the growth and scattering amplitudes satisfy the forward-backward relations~\cite{QKIII,SGPEII}
\EQ{
\label{Gfb}G^{\scriptscriptstyle(-)}(\mbf{u},\mbf{v},\epsilon)=e^{(\epsilon-\mu)/k_BT}G^{\scriptscriptstyle(+)}(\mbf{u},\mbf{v},\epsilon),
}
and 
\EQ{\label{Mfb}
M(\mbf{u},\mbf{v},\epsilon)=e^{-\epsilon/k_BT}M(\mbf{u},\mbf{v},-\epsilon).
}
\subsubsection{Full master equation and its stationary solution}
Defining the condensate band number operator
\EQ{\label{Nc}
N_C\equiv\myint{\mbf{x}}\phi^\dag(\mbf{x})\phi(\mbf{x}),
}
we can write the stationary solution of the full master equation
\EQ{\label{fullmaster}
\dot{\rho}_C=\dot{\rho}_C\big{|}_{\rm Ham}+\dot{\rho}_C\big{|}_{\rm growth}+\dot{\rho}_C\big{|}_{\rm scatt}
}
as
\EQ{\label{rhos}
\rho_{\rm s}\propto\exp{\left(\frac{\mu N_C-H_C}{k_BT}\right)}.
}
This follows from operator identities of the form
\EQ{\fl\label{ident1}
\left\{G^{\scriptscriptstyle (-)}(\mbf{u},\mbf{v},L_C)\phi(\mbf{u}-\mbf{v}/2)\right\}\rho_{\rm s}&=&\rho_{\rm s}\left\{G^{\scriptscriptstyle (+)}(\mbf{u},\mbf{v},L_C)\phi(\mbf{u}-\mbf{v}/2)\right\},\\\fl\label{ident2}
\left\{G^{\scriptscriptstyle (+)}(\mbf{u},\mbf{v},-L_C)\phi^\dag(\mbf{u}-\mbf{v}/2)\right\}\rho_{\rm s}&=&\rho_{\rm s}\left\{G^{\scriptscriptstyle (-)}(\mbf{u},\mbf{v},-L_C)\phi^\dag(\mbf{u}-\mbf{v}/2)\right\},\\\fl\label{ident3}
\left\{M(\mbf{u},\mbf{v},-L_C)U(\mbf{u}-\mbf{v}/2)\right\}\rho_{\rm s}&=&\rho_{\rm s}\left\{M(\mbf{u},\mbf{v},L_C)U(\mbf{u}-\mbf{v}/2)\right\},
}
which are derived using \eref{Gfb}, \eref{Mfb}, \eref{rhos} and the commutators $[\phi(\mbf{x}),N_C]=\phi(\mbf{x})$, and $[U(\mbf{x}),N_C]=0$.
\subsection{Approximate master equation}\label{sec:approxmastereq}
In the high temperature regime the master equation can be linearised to obtain a master equation which can, within approximations that we discuss below, be mapped to a Fokker-Planck equation with positive definite diffusion matrix.
\subsubsection{Approximate treatment of growth}
Since the dependence of the amplitude \eref{Gplus} on $\mbf{v}$ is very sharply peaked around $\mbf{v}=0$, we may treat the growth as a \ital{local} process and approximate the field operators by $\phi(\mbf{u}\pm\mbf{v}/2)\approx \phi(\mbf{u})$. We make the further approximation that
\EQ{
\label{Glinearize1}
G^{\scriptscriptstyle(+)}(\mbf{u},\mbf{v},\epsilon)&\approx& G^{\scriptscriptstyle(+)}(\mbf{u},\mbf{v},0).
}
This represents the fundamental result of the Quantum Kinetic Theory of Bose-Einstein condensation: stimulated growth of the condensate involves collisions between two noncondensate band atoms at right angles such that one of the atoms takes all of the energy of the colliding pair. The remaining atom passes into the condensate. This means that the growth amplitude~\eref{Gplus} can be approximated by its value at zero energy. From the point of view of \cite{SGPEI} this amounts to neglecting the local condensate energy relative to the energy of the collision.
\par
Linearizing the forward-backward relation~\eref{Gfb} then gives
\EQ{\label{Glin2}
G^{(\scriptscriptstyle -)}(\mbf{u},\mbf{v},\epsilon)\approx\left(1-\frac{\mu-\epsilon}{k_BT}\right)G^{(\scriptscriptstyle +)}(\mbf{u},\mbf{v},0)
}
This requires that we are in the high temperature regime in the sense that $(\mu-L_C)\phi(\mbf{x})/k_BT$ is always small. In other words the difference between the chemical potentials of the two bands must be small relative to the temperature. 
\par
Combining the two approximations \eref{Glinearize1} and \eref{Glin2}, we define the growth amplitude 
\EQ{
G(\mbf{x})\equiv\int d^3\mbf{v}\;G^{(\scriptscriptstyle +)}(\mbf{u},\mbf{v},0).
}
The effect of rotation on the growth is merely to dilate the thermal cloud, so that we can obtain the approximate high temperature form by using the approach of~\cite{QKIII} to get
\EQ{\label{approxgrowth}
G(\mbf{x})=\frac{16(k_BT)^2a^3}{\hbar u}\exp{[2(\mu-V_{\rm eff}(\mbf{x}))/k_BT]}
}
where the rotating frame effective potential is defined in~\eref{Veffgrowth}. The result~\eref{approxgrowth} follows immediately from the high temperature calculation of $W^{+}$ in~\cite{QKIII} since the derivation used there is explicitly \ital{local}, so that the chemical potential of the bath can be replaced by the local effective chemical potential of the rotating cloud. In fact, a more accurate form for the growth amplitude that includes the effects of quantum statistics and the cutoff on the noncondensate band distribution, found for the non-rotating case by Davis~\cite{QKPRLII}, can be obtained for the rotating scenario in exactly the same way. Although it is important to include these effects for certain scenarios, in this paper we restrict our attention to the approximate form \eref{approxgrowth}.
\subsubsection{Approximate treatment of scattering}
Expanding the forward-backward relation~\eref{Mfb} to lowest order in $\epsilon/k_BT$ gives
\EQ{
\label{Mlinearize}
M(\mbf{u},\mbf{v},\epsilon)&\approx&\left(1-\frac{\epsilon}{2k_BT}\right)M(\mbf{u},\mbf{v},0).
}
If the condensate band is always reasonably close to equilibrium with the thermal cloud so that the growth linearisation is valid, the approximation for the scattering is the most restrictive. In terms of the eigenoperators of $H_C$ defined by~\cite{QKIII}
\EQ{
[H_C,X_m(\mbf{x})]=-\epsilon_mX_m(\mbf{x}),
}
we have 
\EQ{
L_CX_m(\mbf{x})=\epsilon_m X_m(\mbf{x}),
}
and the condition~\eref{Mlinearize} is a requirement of high temperature in the sense that
\EQ{
|\epsilon_m|\ll 2k_BT,
}
for the eigenvalues of $L_C$ associated with the eigenoperators that contribute significantly to the expansion
\EQ{
\phi(\mbf{x})=\sum_m X_m(\mbf{x}).
}
The highest value of $\epsilon_m$ available is of the order $\;\epsilon_m\approx E_R$, so we require
\EQ{
E_R \ll 2k_BT.
}
In~\cite{QKVI} it was shown that the single particle energy levels of the noncondensate band are very closely approximated by those of the trapping potential, provided the cut-off satisfies
\EQ{
2.5\mu\lesssim E_R.
}
Under these conditions, we require
\EQ{\label{scattlin}
\mu\ll k_BT
}
for the linearisation of the scattering to be valid. 
\subsubsection{Scattering amplitude}
We can compute the amplitude by considering the Fourier transform
\EQ{\fl
\tilde{M}(\mbf{x},\mbf{k},0)&=&\frac{1}{(2\pi)^3}\myint{\mbf{v}}\;e^{-i\mbf{k}\cdot\mbf{v}}M(\mbf{x},\mbf{v},0)\nonumber\\
\fl&&=\frac{2u^2}{(2\pi)^5\hbar^2}\int_{R_{NC}}d^3\mbf{K}_1\int_{R_{NC}}d^3\mbf{K}_2\;F(\mbf{x},\mbf{K}_1)[1+F(\mbf{x},\mbf{K}_2)]\nonumber\\
\fl&&\hspace{3cm}\times\delta(\mbf{K}_1-\mbf{K}_2-\mbf{k})\delta(\omega(\mbf{x},\mbf{K}_1)-\omega(\mbf{x},\mbf{K}_2)),
}
with $\omega(\mbf{x},\mbf{K})$ given by \eref{Krdef}.
\par 
After evaluating the momentum delta function this becomes
\EQ{
\tilde{M}(\mbf{x},\mbf{k},0)&=&\frac{4mu^2}{(2\pi)^5\hbar^3}\int_{R_{NC}}d^3\mbf{K}\;F(\mbf{x},\mbf{K})[1+F(\mbf{x},\mbf{K})]\nonumber\\
&&\hspace{2cm}\times\delta\left(2\left\{\mbf{K}-\frac{m}{\hbar}\mbf{\Omega}\times\mbf{x}\right\}\cdot\mbf{k}-\mbf{k}^2\right).
}
Using the notation $\hbar\omega(\mbf{x},\mbf{K})=e_{\mbf{K}-m\mbf{\Omega}\times\mbf{x}/\hbar}(\mbf{x})$, where
\EQ{\label{edef}
e_{\mbf{K}}(\mbf{x})\equiv\frac{\hbar^2\mbf{K}^2}{2m}+V_{\rm eff}(\mbf{x}),
} 
we can change to the integration variable $\mbf{q}=\mbf{K}-m(\mbf{\Omega}\times\mbf{x})/\hbar$, and obtain
\EQ{\label{Mterm}
\tilde{M}(\mbf{x},\mbf{k},0)&=&\frac{4mu^2}{(2\pi)^5\hbar^3}\int_{R_{NC}}d^3\mbf{q}\;f_{T,\mu}(e_{\mbf{q}}(\mbf{x}))[1+f_{T,\mu}(e_{\mbf{q}}(\mbf{x}))]\nonumber\\
&&\hspace{2cm}\times\delta\left(2\mbf{q}\cdot\mbf{k}-\mbf{k}^2\right),
}
where 
\EQ{
f_{T,\mu}(z)=\frac{1}{e^{(z-\mu)/k_BT}-1}.
}
Note that \eref{Mterm} differs from the corresponding expression (88) of \cite{SGPEII} only in the replacement $V(\mbf{x})\to V_{\rm eff}(\mbf{x})$, and the change of integration region to $e_{\mbf{q}}(\mbf{x})>E_R$.
A straightforward calculation along the lines given in \cite{SGPEII} leads to the form
\EQ{
\tilde{M}(\mbf{x},\mbf{k},0)\approx \frac{16\pi(ak_BT)^2}{\hbar(E_R-\mu)(2\pi)^3|\mbf{k}|}\equiv\frac{{\cal M}}{(2\pi)^3|\mbf{k}|},\label{Mapprox}
}
where the high temperature condition and the fact that the operators upon which $\tilde{M}(\mbf{x},\mbf{k},0)$ acts are restricted to the condensate band have been used to obtain the final form. Note that the ease with which the laboratory frame calculation is shifted to the rotating frame has its origin in the fact that both the distributions and the edge of the integration region only depend on energy. In the semi-classical description we are using the energies in the two frames only differ by a position dependent shift in momentum and a change of the effective potential.
\par
In the spatial variables we find
\EQ{
M(\mbf{x},\mbf{v},0)&=&\frac{\cal M}{(2\pi)^3}\myint{\mbf{k}}\frac{e^{i\mbf{k}\cdot\mbf{v}}}{|\mbf{k}|},
}
and the action of the scattering operator is
\EQ{
\myint{\mbf{x}^\prime}M\left(\frac{\mbf{x}+\mbf{x}^\prime}{2},\mbf{x}-\mbf{x}^\prime,0\right)f(\mbf{x}^\prime)=\frac{\cal M}{\sqrt{-\nabla^2}}f(\mbf{x}).
}
It is clear that the scattering term is entirely non-local. 
\subsubsection{The Master equation}
Using the above approximations leads to the Master equation
\EQ{\label{appmaster}
\dot{\rho}_C=\dot{\rho}_C\big{|}_{\rm Ham}+\dot{\rho}_C\big{|}_{\rm growth}+\dot{\rho}_C\big{|}_{\rm scatt}
}
where the Hamiltonian evolution is given by~\eref{Hamrho} and the growth and scattering terms are
\EQ{\fl
\dot{\rho}_C\big{|}_{\rm growth}&=&\;\;\;\;\int d^3\mbf{x}\;G(\mbf{x})\Big\{\left[[\phi(\mbf{x}),\rho_C],\phi^\dag(\mbf{x})\right]+\left[\phi(\mbf{x}),[\rho_C,\phi^\dag(\mbf{x})]\right]\Big\}\nonumber\\\fl
&&\label{Gmasterlin}-\int d^3\mbf{x}\;\frac{G(\mbf{x})}{k_BT}\Big\{\left[\left\{(\mu-L_C)\phi(\mbf{x})\right\}\rho_C,\phi^\dag(\mbf{x})\right]\nonumber\\\fl 
&&\;\;\;\;\;\;\;\;\;\;\;\;\;\;\;\;\;\;\;\;\;\;+\left[\phi(\mbf{x}),\rho_C\left\{(\mu+L_C)\phi^\dag(\mbf{x})\right\}\right]\Big\}\;\;\;\;\;\;\;\;\;
}
and
\EQ{\fl\label{scattmasterlin}
\dot{\rho}_C\big{|}_{\rm scatt}&=&-\int d^3\mbf{u}\int d^3\mbf{v}\;M(\mbf{u},\mbf{v},0)\Big\{
\left[U(\mbf{u}+\mbf{v}/2),[U(\mbf{u}-\mbf{v}/2),\rho_C]\;\right]\nonumber\\\fl
&&\;\;\;\;\;\;\;\;\;\;\;\;\;\;\;\;\;\;\;\;\;\;\;\;\;\;\;\;\;\;+\frac{1}{2k_BT}\left[U(\mbf{u}+\mbf{v}/2),[(L_CU(\mbf{u}-\mbf{v}/2)),\rho_C]_{\scriptscriptstyle+}\right]\Big\},\;\;\;\;\;\;\;\;
}
and where $[\;,\;]_{\scriptscriptstyle +}$ denotes the anti-commutator. The scattering terms are a non-local form of the quantum Brownian motion master equation~\cite{QN}.
\subsubsection{Non-Lindblad terms}
The Lindblad property is a highly desirable property for Markovian master equations since it guarantees that the positive semi-definiteness of density operators is preserved by the dissipation.
The scattering terms in the master equation, given by~\eref{scattmasterlin}, cannot be cast in Lindblad form. 
There are also growth terms which are not of Lindblad form, arising from the kinetic and interaction terms in the second line of~\eref{Gmasterlin}. The kinetic terms take the form
\EQ{\fl\label{Lindbladkin}
\dot{\rho}_C\Big{|}_{\rm kin}&=&\int d^3\mbf{x}\;\frac{G(\mbf{x})}{k_BT}\frac{\hbar^2}{2m}\left[2\nabla\phi(\mbf{x})\rho_C\nabla\phi^\dag(\mbf{x})-\nabla\phi^\dag(\mbf{x})\nabla\phi(\mbf{x})\rho_C-\rho_C\nabla\phi^\dag(\mbf{x})\nabla\phi(\mbf{x})\right]\nonumber\\\fl
&&+\int d^3\mbf{x}\;\frac{\hbar^2}{2mk_BT}\nabla G(\mbf{x})\cdot\big[\phi(\mbf{x})\rho_C\nabla\phi^\dag(\mbf{x})+\nabla\phi(\mbf{x})\rho_C\phi^\dag(\mbf{x})\nonumber\\\fl
&&-\rho_C\nabla\phi^\dag(\mbf{x})\phi(\mbf{x})-\phi^\dag(\mbf{x})\nabla\phi(\mbf{x})\rho_C\big].
}
If the thermal cloud varies slowly over the condensate region we may neglect $\nabla G(\mbf{x})$ and recover a dissipative kinetic term in Lindblad form given by the first line of~\eref{Lindbladkin}.
\par
The interaction terms in~\eref{Gmasterlin} are
\EQ{\fl\label{nonlindbladint}
\dot{\rho}_C\Big{|}_{\rm int}&=&\int d^3\mbf{x}\;\frac{G(\mbf{x})}{k_BT}\;u\Big([\phi^\dag(\mbf{x})\phi(\mbf{x})\phi(\mbf{x})\rho_C,\phi^\dag(\mbf{x})]+[\phi(\mbf{x}),\rho_C\phi^\dag(\mbf{x})\phi^\dag(\mbf{x})\phi(\mbf{x})]\Big),
}
which is not of Lindblad form. However, Munro and Gardiner~\cite{Munro1996} have shown that the effect of non-Lindblad dissipation is usually to generate small initial transients in the time evolution, so we may be confident that the dissipative evolution will still give physically meaningful predictions. 
\section{The Wigner phase-space representation}\label{subsec:TWR}
\subsection{Wigner distribution}
We expand the projected field operator in the single particle basis
\EQ{
\phi(\mathbf{x})=\sum_n^- a_nY_n(\mathbf{x})
}
where the bar notation indicates that the summation is carried out up to
the energy cut-off of the condensate band, and 
\EQ{
\left[a_n,a_m^\dag\right]=\delta_{nm},\;\;\left[a_n^\dag,a_m^\dag\right]=\left[a_n^\dag,a_m^\dag\right]=0.
} 
Following~\cite{QN}, we may define the symmetrically ordered quantum characteristic function 
\EQ{
\chi_{\scriptscriptstyle W}(\{\lambda_n,\lambda^*_n\})&\equiv&\tr{
\left\{\rho_C\;\exp{\left(\sum_m^-\lambda_m a^\dag_m-\lambda^*_m a_m\right)}\right\}}.
}
The multimode Wigner function is then given by
\EQ{\fl
W(\left\{\alpha_n,\alpha^*_n\right\})&=&\int\prod_m^- \frac{d^2\lambda_m}{\pi^2}
\exp{\left(\sum_p^-\lambda^*_p
\alpha_p-\lambda_p\alpha_p^*\right)}\chi_{\scriptscriptstyle W}(\{\lambda_q,\lambda^*_q\}).
}
Moments of the Wigner distribution give symmetrically ordered operator averages, for example
\EQ{\label{wigav1}
\int \prod_n^-d^2\alpha_n |\alpha_q|^2W(\left\{\alpha_m,\alpha^*_m\right\})=\left\langle\frac{a_q^\dag a_q+a_qa_q^\dag}{2}\right\rangle.
}
\subsection{Operator correspondences}
Defining the projected stochastic field $\alpha(\mbf{x})$ and the projected derivative operators as
\EQ{\label{alphadef}
\alpha(\mathbf{x})&\equiv&\sum_n^- \alpha_nY_n(\mathbf{x}),\\
\DDP{\alpha(\mathbf{x})}&\equiv&\sum_n^-
Y_n^*(\mathbf{x})\frac{\partial}{\partial \alpha_n},\\
\DDP{\alpha^*(\mathbf{x})}&\equiv&\sum_n^-
Y_n(\mathbf{x})\frac{\partial}{\partial \alpha_n^*}
}
one then finds functional operator correspondences for the Wigner function~\cite{QN}
\EQ{\label{opcorr1}
\phi(\mathbf{x})\rho&\longleftrightarrow&\left(\alpha(\mathbf{x})+\frac{1}{2}\frac{\bar{\delta}}{\bar{\delta}\alpha^*(\mathbf{x})}\right)W,\\
\label{opcorr2}\phi^\dag(\mathbf{x})\rho&\longleftrightarrow&\left(\alpha^*(\mathbf{x})-\frac{1}{2}\frac{\bar{\delta}}{\bar{\delta}\alpha(\mathbf{x})}\right)W,\\
\label{opcorr3}\rho \phi(\mathbf{x})&\longleftrightarrow&\left(\alpha(\mathbf{x})-\frac{1}{2}\frac{\bar{\delta}}{\bar{\delta}\alpha^*(\mathbf{x})}\right)W,\\
\label{opcorr4}\rho \phi^\dag(\mathbf{x})&\longleftrightarrow&\left(\alpha^*(\mathbf{x})+\frac{1}{2}\frac{\bar{\delta}}{\bar{\delta}\alpha(\mathbf{x})}\right)W,
}
which are used to map the master equation~\eref{appmaster} to an equation of motion for $W$.
\par
The field average corresponding to~\eref{wigav1} is
\EQ{\label{uvdivergent}
\prod_n^-\int d^2\alpha_n\;|\alpha(\mbf{x})|^2W(\left\{\alpha_m,\alpha^*_m\right\})&=&\left\langle\frac{\phi^\dag(\mbf{x})\phi(\mbf{x})+\phi(\mbf{x})\phi^\dag(\mbf{x})}{2}\right\rangle\nonumber\\ &=&\left\langle\phi^\dag(\mbf{x})\phi(\mbf{x})\right\rangle+\frac{\delta_C(\mbf{x},\mbf{x})}{2},
}
where the \ital{condensate band delta function}
\EQ{\label{deltac}
\delta_C(\mbf{x},\mbf{x}^\prime)\equiv\sum_n^-Y_n(\mbf{x})Y^*_n(\mbf{x}^\prime)=[\phi(\mbf{x}),\phi^\dag(\mbf{x}^\prime)]
}
is always well defined at $\mbf{x}=\mbf{x}^\prime$ for a finite cut-off energy. The delta-function contribution in \eref{uvdivergent} exhibits a well known issue with using the Wigner representation of quantum field theory. The theory is ultraviolet divergent, a feature which is absent from the positive-P formulation~\cite{Steel1998}. Nevertheless, the Wigner representation has some utility when treating ultra-cold Bose gases subject to the cold-collision interaction Hamiltonian \eref{ccint}. Within the Wigner representation there is an approximation available which is known as the {\em classical field method} or {\em truncated Wigner method}.
\subsection{Truncated Wigner method}\label{truncatedWig}
By way of the operator correspondences \eref{opcorr1} -- \eref{opcorr4}, the Hamiltonian terms~\eref{H0def1} generate the time evolution
\EQ{\fl\label{HCthirdorder}
\frac{\partial W}{\partial t}\Big{|}_{\rm Ham}&=&\myint{\mbf{x}}\Bigg\{\nonumber\\\fl
&&\frac{i}{\hbar}\DDP{\alpha(\mbf{x})}\Big(-\frac{\hbar^2\nabla^2}{2m}+i\hbar\mbf{\Omega}\cdot(\mbf{x}\times\nabla)+V(\mbf{x})+u[|\alpha(\mbf{x})|^2-\delta_C(\mbf{x},\mbf{x})]\Big)\alpha(\mbf{x})\nonumber\\\fl
&-&\frac{i}{\hbar}\DDP{\alpha^*(\mbf{x})}\Big(-\frac{\hbar^2\nabla^2}{2m}-i\hbar\mbf{\Omega}\cdot(\mbf{x}\times\nabla)+V(\mbf{x})+u[|\alpha(\mbf{x})|^2-\delta_C(\mbf{x},\mbf{x})]\Big)\alpha^*(\mbf{x})\nonumber\\\fl
&&+\frac{iu}{4\hbar}\DPDP{\alpha(\mbf{x})}{\alpha^*(\mbf{x})}\left(\alpha^*(\mbf{x})\DDP{\alpha^*(\mbf{x})}-\alpha(\mbf{x})\DDP{\alpha(\mbf{x})}\right)\Bigg\}W.
}
 The third order terms present a significant difficulty in the sense that there is no equivalent diffusion process which would admit a formulation in terms of stochastic differential equations~\cite{QN,SM}. 
Although stochastic \ital{difference} equations can be found which are formally equivalent to the generalized Fokker-Planck equation~\cite{Plimak2001}, they are difficult to use. The \ital{truncated Wigner method}~\cite{Kinsler1991,Steel1998,QO,Sinatra2000,Sinatra2001,Sinatra2002}, is based on the observation that the third order derivatives are small when the modes of interest are highly occupied throughout. If we neglect the third order terms in \eref{HCthirdorder}, the resulting equation of motion reduces to a Liouville equation since it lacks any diffusion terms, and the equation of motion for the classical field $\alpha(\mbf{x})$ has the superficial appearance of the Gross-Pitaevskii equation. It should be stressed that there are two important differences. (i) As we have seen, moments of the Wigner function give symmetrically ordered operator averages which must be calculated over a large number of trajectories. (ii) The initial state must also be chosen according to the Wigner transform of the initial density matrix. Although the equation takes the form of the Gross-Pitaevskii equation, only in the limit that the initial Wigner function is a Dirac delta distribution
do we recover the Gross-Pitaevskii theory. We note Polkovnikov has shown that truncated Wigner approximation arises naturally as the first in a hierarchy of quantum corrections to the classical field evolution described by the Gross-Pitaevskii equation~\cite{Polkovnikov2003b,Polkovnikov2003}. 

The application of the truncated Wigner approximation to the generalized Fokker-Planck equation arising from~\eref{appmaster} leads to a genuine Fokker-Planck equation with a positive definite diffusion matrix which can then be mapped to a stochastic differential equation of motion for the condensate band. 

\section{Condensate band Fokker-Planck equation}
We can now use standard methods~\cite{QN} to map the Master equation to a genuine Fokker-Planck equation for the Wigner distribution. 
\subsection{Hamiltonian terms}
Defining the Gross-Pitaevskii Hamiltonian corresponding to~\eref{H0def1}
\EQ{
H_{\rm GP}&=&\int d^3\mbf{x}\;\alpha^*(\mbf{x})\left(-\frac{\hbar^2\nabla^2}{2m}+i\hbar\mbf{\Omega}\cdot (\mbf{x}\times\nabla)+V(\mbf{x})\right)\alpha(\mbf{x})\nonumber\\\label{GPHdef}
&&+\frac{u}{2}\int d^3\mbf{x}\;\alpha^*(\mbf{x})\alpha^*(\mbf{x})\alpha(\mbf{x})\alpha(\mbf{x}),
}
the GP operator $L_{\rm GP}$
\EQ{\label{LGPdef}
\PP L_{\rm GP}\alpha(\mbf{x})\equiv\DP{H_{\rm GP}}{\alpha^*(\mbf{x})},
}
and neglecting third order derivatives in~\eref{HCthirdorder}, we can express the Hamiltonian term as
\EQ{\label{HamW}
\frac{\partial W}{\partial t}\Big{|}_{\rm Ham}&=&\int d^3\mbf{x}\; \left\{-\DDP{\alpha(\mbf{x})}\left(-\frac{i}{\hbar} L_{\rm GP}\alpha(\mbf{x})\right)+{\rm c.c.}\right\}W.\;\;\;\;\;\;\;
}
Note that $L_{\rm GP}$ is implicitly projected since the action of a projected functional derivative is equivalent to non-projected functional differentiation of a projected operator -- a consequence of the Hermiticity of projectors. We can express \eref{HamW} as
\EQ{
\frac{\partial W}{\partial t}\Big{|}_{\rm Ham}&=&\int d^3\mbf{x}\; \left\{-\frac{\delta}{\delta\alpha(\mbf{x})}\left(-\frac{i}{\hbar} \PP L_{\rm GP}\alpha(\mbf{x})\right)+{\rm c.c.}\right\}W,\;\;\;\;\;\;\;
}
from which can use the usual correspondence rules of functional Fokker-Planck equations~\cite{Steel1998} to identify the drift term as the projected Gross-Pitaevskii equation of motion~\cite{Davis2001b} for $\alpha(\mbf{x},t)$, as seen in the first line of \eref{rsgpefullform}.
\subsection{Growth terms}
The master equation terms~\eref{Gmasterlin} give
\EQ{\fl
\frac{\partial W}{\partial t}\Big{|}_{\rm growth}&=&\int d^3\mbf{x}\; G(\mbf{x})\Bigg\{-\DDP{\alpha(\mbf{x})}\left(\frac{(\mu-L_{\rm GP})\alpha(\mbf{x})}{k_BT}\right)+{\rm c.c.}\nonumber\\\fl
&&+\DDP{\alpha^*(\mbf{x})}\DDP{\alpha(\mbf{x})}+{\rm c.c.}\Bigg\}W
}
which contributes a drift term and a thermal noise term to the stochastic Gross-Pitaevskii equation.
\subsection{Scattering terms}
Our development of the formalism thus far has largely followed~\cite{SGPEII}, the only modification being the $-\mbf{\Omega}\cdot \mbf{L}$ term in the single particle Hamiltonian arising from the rotation. Similarly, for the scattering there are extra terms arising from the formulation in a rotating reference frame. 
\par
To derive the scattering terms in the Fokker-Planck equation we need to evaluate~\eref{scattmasterlin}. It is easily seen that
\EQ{
L_C U(\mbf{x})=-i\hbar\nabla\cdot \mbf{j}_C(\mbf{x})
}
where the \ital{condensate band current} operator
\EQ{
\mbf{j}_C(\mbf{x})\equiv \frac{i\hbar}{2m}\left([\nabla\phi^\dag(\mbf{x})]\phi(\mbf{x})-\phi^\dag(\mbf{x})\nabla\phi(\mbf{x})\right)-(\mbf{\Omega}\times\mbf{x})\phi^\dag(\mbf{x})\phi(\mbf{x})
}
exhibits a centrifugal term in the rotating frame. The corresponding Gross-Pitaevskii current
\EQ{\label{JGPdef}
\mbf{j}_{\rm GP}(\mbf{x})\equiv\frac{i\hbar}{2m}[\alpha(\mbf{x})\nabla\alpha^*(\mbf{x})-\alpha^*(\mbf{x})\nabla\alpha(\mbf{x})]-(\mbf{\Omega}\times\mbf{x})\alpha^*(\mbf{x})\alpha(\mbf{x})
}
arises in the drift vector of the Fokker-Planck equation, which takes the form
\EQ{\fl\label{scattFPE}
\frac{\partial W}{\partial t}\Big{|}_{\rm scatt}&=&\int d^3\mbf{x}\int d^3\mbf{x}^\prime\; M\left(\frac{\mbf{x}+\mbf{x}^\prime}{2},\mbf{x}^\prime-\mbf{x},0\right)\nonumber\\\fl\nonumber
&&\;\;\;\;\;\;\;\;\;\;\;\;\;\;\;\;\;\;\times\Bigg\{
-\DDP{\alpha(\mbf{x}^\prime)}\alpha(\mbf{x}^\prime)\frac{i\hbar\nabla\cdot \mbf{j}_{\rm GP}(\mbf{x})}{k_BT}+{\rm c.c.}\;\;\;\;\;\;\;\;\;\;\;\;\;\;\\\fl
&&\;\;\;\;\;\;\;\;\;\;\;\;\;\;\;\;\;\;\;\;\;\;\;\;\;\;+\DDP{\alpha(\mbf{x}^\prime)}\alpha(\mbf{x}^\prime)\DDP{\alpha^*(\mbf{x})}\alpha^*(\mbf{x})+{\rm c.c.}\nonumber\\\fl
&&\;\;\;\;\;\;\;\;\;\;\;\;\;\;\;\;\;\;\;\;\;\;\;\;\;\;-\DDP{\alpha(\mbf{x}^\prime)}\alpha(\mbf{x}^\prime)\DDP{\alpha(\mbf{x})}\alpha(\mbf{x})+{\rm c.c.}\Bigg\}W.
}
\subsection{Full Fokker-Planck equation and its stationary solution}
The Wigner distribution evolves according to the Fokker-Planck equation 
\EQ{
\frac{\partial W}{\partial t}=\frac{\partial W}{\partial t}\Big{|}_{\rm Ham}+\frac{\partial W}{\partial t}\Big{|}_{\rm growth}+\frac{\partial W}{\partial t}\Big{|}_{\rm scatt}.
}
By defining the GP number operator corresponding to \eref{Nc}
\EQ{\label{Ngp}
N_{\rm GP}\equiv\myint{\mbf{x}}\alpha^*(\mbf{x})\alpha(\mbf{x}),
}
the equation of motion can be written as
\EQ{\fl
\frac{\partial W}{\partial t}&=&\myint{\mbf{x}}\DDP{\alpha(\mbf{x})}\left\{\frac{iW}{\hbar}\DP{H_{\rm GP}}{\alpha^*(\mbf{x})}+G(\mbf{x})\left[\DP{W}{\alpha^*(\mbf{x})}-\frac{W}{k_BT}\DP{(\mu N_{\rm GP}-H_{\rm GP})}{\alpha^*(\mbf{x})}\right]\right\}\nonumber\\\fl
\nonumber\\\fl
&&-\myint{\mbf{x}}\DDP{\alpha^*(\mbf{x})}\left\{\frac{iW}{\hbar}\DP{H_{\rm GP}}{\alpha(\mbf{x})}-G(\mbf{x})\left[\DP{W}{\alpha(\mbf{x})}-\frac{W}{k_BT}\DP{(\mu N_{\rm GP}-H_{\rm GP})}{\alpha(\mbf{x})}\right]\right\}\nonumber\\\fl
\nonumber\\\fl
&&+\int d^3\mbf{x}\myint{x^\prime}M\left(\frac{\mbf{x}+\mbf{x}^\prime}{2},\mbf{x}-\mbf{x}^\prime,0\right)\left[\alpha(\mbf{x})\DDP{\alpha(\mbf{x})}-\alpha^*(\mbf{x})\DDP{\alpha^*(\mbf{x})}\right]\nonumber\\\fl
\nonumber\\\fl
&&\times\left\{\alpha^*(\mbf{x}^\prime)\left[\DP{W}{\alpha^*(\mbf{x}^\prime)}+\frac{W}{k_BT}\DP{H_{\rm GP}}{\alpha^*(\mbf{x}^\prime)}\right]-\alpha(\mbf{x}^\prime)\left[\DP{W}{\alpha(\mbf{x}^\prime)}+\frac{W}{k_BT}\DP{H_{\rm GP}}{\alpha(\mbf{x}^\prime)}\right]\right\}.
}
Irrespective of the forms of $G(\mbf{x})$ and $M(\mbf{u},\mbf{v},0)$, the Fokker-Planck equation has the grand canonical stationary solution
\EQ{
W\propto \exp{\left(\frac{\mu N_{\rm GP}-H_{\rm GP}}{k_BT}\right)}.
}
\section{Rotating stochastic Gross-Pitaevskii equation}\label{sec:rsgpe}
\subsection{Full form of the SGPE}
In the rotating frame the full non-local form is given by the stochastic differential equation in Stratonovich form
\EQ{\fl
(S)d\alpha(\mbf{x},t)&=&-\frac{i}{\hbar}\PP L_{\rm GP}\alpha(\mbf{x})dt\nonumber\\\fl
&&
\PP\left\{\frac{G(\mbf{x})}{k_BT}(\mu-L_{\rm GP})\alpha(\mbf{x})dt+dW_G(\mbf{x},t)\right\}\nonumber
\\\fl
&&+\PP\Bigg\{\myint{\mbf{x}^\prime}M\left(\frac{\mbf{x}^\prime+\mbf{x}}{2},\mbf{x}-\mbf{x}^\prime,0\right)\frac{ i\hbar\alpha(\mbf{x})}{k_BT} \nabla\cdot \mbf{j}_{\rm GP}(\mbf{x}^\prime)dt\;\;\;\;\;\;\;\;\;\;\;\;\;\;\;\nonumber\\\fl\label{rsgpefullform}
&&\;\;\;\;\;\;\;\;\;\;\;\;\;+idW_{M}(\mbf{x},t)\alpha(\mbf{x})\Bigg\},
}
where the growth noise is complex, the scattering noise is real, they are independent of each other, and satisfy
\EQ{
dW_G^*(\mbf{x},t)dW_G(\mbf{x}^\prime,t)&=&2G(\mbf{x})\delta_C(\mbf{x},\mbf{x}^\prime)dt,\\
dW_G(\mbf{x},t)dW_G(\mbf{x}^\prime,t)&=&dW_G^*(\mbf{x},t)dW_G^*(\mbf{x}^\prime,t)=0\\
dW_{M}(\mbf{x},t)dW_{ M}(\mbf{x}^\prime,t)&=&2M\left(\frac{\mbf{x}^\prime+\mbf{x}}{2},\mbf{x}-\mbf{x}^\prime,0\right) dt.
}
\subsection{Approximate non-local form}
In the rotating frame the approximate non-local form of the SGPE~\cite{SGPEII}, which is probably the most practical form for realistic simulations, is given by the Stratonovich SDE
\EQ{\fl
(S)d\alpha(\mbf{x},t)&=&-\frac{i}{\hbar}\PP L_{\rm GP}\alpha(\mbf{x})dt\nonumber\\\fl
&&+\PP\Big\{\frac{G(\mbf{x})}{k_BT}(\mu-L_{\rm GP})\alpha(\mbf{x})dt+dW_G(\mbf{x},t)\Big\}\nonumber
\\\fl\label{SGPEaloc}
&&+\PP\Big\{\frac{\hbar{\cal M }i\alpha(\mbf{x})}{k_BT}\frac{1}{\sqrt{-\nabla^2}} \nabla\cdot \mbf{j}_{\rm GP}(\mbf{x})dt+idW_{{\cal M}}(\mbf{x},t)\alpha(\mbf{x})\Big\},
}
where the growth noise is complex, the scattering noise is real, they are independent of each other, and satisfy
\EQ{\label{Gnoisecorr1}
dW_G^*(\mbf{x},t)dW_G(\mbf{x}^\prime,t)&=&2G(\mbf{x})\delta_C(\mbf{x},\mbf{x}^\prime)dt,\\
\label{Gnoisecorr2}dW_G(\mbf{x},t)dW_G(\mbf{x}^\prime,t)&=&dW_G^*(\mbf{x},t)dW_G^*(\mbf{x}^\prime,t)=0\\
\label{Gnoisecorr3}dW_{{\cal M}}(\mbf{x},t)dW_{{\cal M}}(\mbf{x}^\prime,t)&=&\frac{2{\cal M}}{\sqrt{-\nabla^2}}\delta_C(\mbf{x},\mbf{x}^\prime)dt.
}
There are some corrections to mention here in the scattering terms of the forms \eref{rsgpefullform} and \eref{SGPEaloc}, compared to equations (118) and (122) of~\cite{SGPEII}. In particular, we note that the scattering term in \cite{SGPEII}, proportional to $\alpha^*(\mbf{x},t)\bar{L}_C\alpha(\mbf{x},t)-\alpha(\mbf{x},t)(\bar{L}_C\alpha(\mbf{x},t))^*$, is subtly different to the corresponding expression in \eref{rsgpefullform}, \eref{SGPEaloc}, $\nabla \cdot \mbf{j}_{\rm GP}$, since there are no projectors in the latter. Furthermore, the form of ${\cal M}$ arising in our \eref{Mapprox} and \eref{SGPEaloc} is bigger than the form defined in equation (97) of \cite{SGPEII} by a factor of $(2\pi)^3$.
\section{Theory of irreversible vortex nucleation}\label{sec:svge}
Now that we have developed a theory that includes the effects of a rotating thermal cloud on a trapped BEC, we will examine the connection between a simplified form of the theory described in \sref{sec:rsgpe} and the other approaches that have been used to investigate the dynamics of vortex lattice formation.
\par
Much effort has been directed at understanding the process of vortex lattice formation in trapped Bose-Einstein condensates~\cite{Tsubota2002,Penckwitt2002,Williams2002b,Lobo2003}. It is well known that the minimum energy state of a BEC in a rotating reference frame usually contains vortices\footnote[1]{In the regime of slow rotation there are states that have nontrivial phase variation but are nevertheless vortex free~\cite{APThesis}.}. An obvious consequence of this fact is that there must be some irreversible dynamics for the system to proceed from a laboratory frame ground state to a rotating frame ground state; physically this evolution arises from either (i) contact of the condensate with a rotating thermal cloud, or (ii) turbulent motion arising from dynamical instability. Mechanism (ii) has been already been the subject of significant research~\cite{Madison2001,Sinha2001,Kasamatsu2003,Lobo2003}. In this paper we are concerned with (i).
\par 
Although much work has been carried out using phenomenological and hydrodynamic theories~\cite{Tsubota2002,Penckwitt2002,Williams2002b}, the precise description of the thermodynamic scenario has, until now, been an oustanding unsolved problem. Furthermore, it is to be expected that the combination of dynamical instability and nonlinear interactions \ital{generates} a thermal component, which in turn acts as a rotating reservoir that imparts angular momentum to the condensate. We therefore wish to formulate a precise description of a trapped condensate in contact with a rotating thermal cloud, since it is this aspect of the theory that provides the irreversibility required for vortex lattice formation. As is often the case in the theory of dilute gases, we are able to develop a first principles formulation of the theory that goes beyond phenomenology.
\par
The recent history of the phenomenology of damped condensates can be traced to the work of Choi, Morgan and Burnett~\cite{Choi1998} who proposed that dissipation may be incorporated into the GPE theory by introducing a small imaginary component to the time evolution. While this approach has been very successful in practice, the theory is entirely phenomenological. The outstanding issue we have resolved is the consistent description of a rotating thermal reservoir, including the noise arising from Bose-enhanced scattering into the condensate band. We will now show that a simplified form of the SGPE theory leads in a natural way to a picture of GPE evolution in `complex time', as is to be expected on physical grounds for a system coupled to a thermal reservoir. The damped GPE approach of~\cite{Choi1998,Tsubota2002,Kasamatsu2003,Penckwitt2002,SGPEII,Williams2002b} is very close to this kind of description; here we make the connection rigorous.
\subsection{Vortex growth equation}
An equation of motion which describes the dissipative dynamics of vortex lattice formation can be obtained from \eref{SGPEaloc} with some additional approximations. The scattering terms are neglected on the grounds that they primarily lead to a renormalisation of the effective growth rate~\cite{Anglin1999,QKVI}, and so can usually be regarded as a second order effect. We make the simplifying approximation that the thermal cloud is spatially smooth over the region of the condensate, and then set $dW_G\to0$, $\alpha\to \psi$, $\hbar G(\mbf{x})/k_BT\to \gamma$, and $\PP\to 1$, to find

\EQ{\label{VGE}
i\hbar\frac{\partial \psi(\mbf{x},t)}{\partial t}=L_{\rm GP}\psi(\mbf{x},t)+i\gamma(\mu-L_{\rm GP})\psi(\mbf{x},t).
}
This is the simplest description of a condensate in contact with a noiseless rotating thermal cloud at fixed 
chemical potential $\mu$. We make the further transformation $\psi=\alpha e^{-i\mu t/\hbar}$, to obtain
\EQ{\label{complextimevge}
i\hbar\frac{\partial \alpha(\mbf{x},t)}{\partial t}= (1-i\gamma)(L_{\rm GP}-\mu)\alpha(\mbf{x},t),
} 
and it is clear that the condensate wavefunction is evolving in `complex time' obtained from the standard GPE evolution by the replacement $dt \to dt(1-i\gamma)$.
From \eref{GPHdef} and \eref{Ngp} it is easy to show that 
\EQ{\label{monodecrease}
\frac{\partial (H_{\rm GP}-\mu N_{\rm GP})}{\partial t}=-\frac{2\gamma}{\hbar}\myint{\mbf{x}}|(\mu-L_{\rm GP})\alpha(\mbf{x},t)|^2,
}
which shows that the evolution minimises $H_{\rm GP}-\mu N_{\rm GP}$ and the wavefunction evolves into the rotating frame ground state of the GPE with the same chemical potential as the thermal cloud -- a vortex lattice\footnote[1]{For simplicity we have set $\PP\to 1$. However, note that \eref{monodecrease} still holds if we relax this condition~\cite{Bradley2005b}. In other words the evolution toward the  ground state expressed by \eref{monodecrease} also occurs when the theory explicitly includes a high energy cutoff separating condensate and thermal cloud.}. 
\par
Starting from first principles we have arrived at a satisfactory theory of the dissipative dynamics of vortex lattice formation due to energetic (thermodynamic) instability -- {\em without thermal noise}. In essence the theory is obtained by shifting to the cloud frame, and then carefully accounting for interactions between cloud and condensate. The principal aim of this paper is to provide a theory which also accounts for the physical role of thermal noise in this process.
We now briefly outline the differences between this and previous formulations.

\subsection{Phenomenological and hydrodynamic models of dissipation}
The numerical vortex nucleation studies of Tsubota \etal~\cite{Tsubota2002}, and later Kasamatsu \etal~\cite{Kasamatsu2003}, were based upon a phenomenological damped Gross-Pitaevskii equation. This description was introduced in \cite{Tsubota2002} on the grounds that it gives the correct equilibrium solution, and that it follows naturally from the damped GPE approach of~\cite{Choi1998}. 
\par
The {\em physical argument} for such an equation, was first provided by Gardiner \etal~\cite{SGPEI}, and the equation of motion, found via reasoning we will explain below, was subsequently used by Penckwitt~\etal~\cite{Penckwitt2002} to investigate vortex lattice formation. Like the Tsubota equation, the derivation involved a hydrodynamic approximation, in particular it made use of the local energy concept and the approximation $\mu_{\rm c}(\mbf{x},t)\psi\approx i\hbar \partial \psi/\partial t$, where $\mu_{\rm c}(\mbf{x},t)$ denotes the local condensate chemical potential. As described later in~\cite{Kasamatsu2003}, the equation of motion may also be obtained via similar reasoning from the generalized finite temperature GPE derived by Zaremba \etal (ZNG)~\cite{Zaremba1999}; we briefly reiterate the argument here. 
\par 
In the rotating frame the ZNG equation takes the form
\EQ{\label{ZNG}\fl
i\hbar\frac{\partial \psi(\mbf{x},t)}{\partial t}=\left(-\frac{\hbar^2\nabla^2}{2m}+V_{\rm trap}(\mbf{x})-\Omega L_z+u|\psi(\mbf{x},t)|^2+2u\tilde{n}(\mbf{x},t)-i\Gamma\right)\psi(\mbf{x},t),
}
where $\tilde{n}(\mbf{x},t)$ is the noncondensate density, and $\Gamma$ describes collisions with the noncondensate. This equation of motion has been derived within a hydrodynamic approximation, utilising the notion of \ital{local} energy attributed to a smoothly varying condensate wavefunction. We further neglect the noncondensate density under the assumption $|\psi|^2\gg\tilde{n}$, and note that under a local equilibrium distribution for the thermal cloud the growth rate is proportional to the difference between the local chemical potentials of the condensate and noncondensate $\Gamma\propto \mu_{\rm nc}(\mbf{x},t)-\mu_{\rm c}(\mbf{x},t)$; the constant of proportionality will be denoted by $\gamma$. 
\par 
We now make the crucial approximation that generates the dissipative GPE evolution, by putting $\mu_{\rm c}(\mbf{x},t)\psi\approx -i\hbar(\partial \psi/\partial t)$. To reach the final form, one then proceeds by neglecting the space and time dependence of the noncondensate chemical potential ($\mu_{\rm nc}(\mbf{x},t)\to \mu_{\rm nc}=\mu$), and making the further transformation $\psi\to \psi e^{-i\mu t/\hbar}$. We finally arrive at the equation of motion introduced by Tsubota \etal~\cite{Tsubota2002}
\EQ{\label{tsubota}\fl
(i-\gamma)\hbar\frac{\partial \psi(\mbf{x},t)}{\partial t}=\left(-\frac{\hbar^2\nabla^2}{2m}+V_{\rm trap}(\mbf{x})-\Omega L_z+u|\psi(\mbf{x},t)|^2-\mu\right)\psi(\mbf{x},t).
}
Casting this in the form
\EQ{
i\hbar\frac{\partial \psi(\mbf{x},t)}{\partial t}= \frac{1-i\gamma}{1+\gamma^2}(L_{\rm GP}-\mu)\psi(\mbf{x},t),
}
we see that in the small $\gamma$ limit we recover the vortex growth equation \eref{complextimevge}. However, for non-negligible $\gamma$ the imaginary {\em and} real time dynamics generated by \eref{tsubota} are slower than those of the vortex growth equation \eref{VGE}, a consequence of shifting to complex time without accounting for the change in the length of the time vector in the complex plane.
\par 
We stress that neither of these approximations have been made in obtaining the SGPE \eref{SGPEaloc}, and the simpler vortex growth equation \eref{VGE}. Furthermore, the simple growth equation has all the necessary physical properties for the correct damped GPE description of vortex lattice formation.
\section{Conclusion and Outlook}
We have extended the SGPE theory developed in references \cite{SGPEI,SGPEII} to include the possibility that the thermal cloud may be in a state of rotation -- a situation of considerable experimental interest. The theory is suitable for treating the experimentally interesting regime of {\em rotating Bose-Einstein condensation}~\cite{Haljan2001}, where one may expect that vortices can form during condensation, effectively being `pinned' in the condensate so that subsequent scattering into vortex states becomes Bose-enhanced. This is an entirely different physical regime to the coherent vortex formation mechanism arising from stirring a pre-existing condensate~\cite{Madison2000}.
\par
We have examined the connection between the SGPE theory derived here and the phenomenological and hydrodynamic approaches that have been used thus far to model vortex lattice formation. We have shown that a simplified vortex growth equation has all the desirable properties of the other mean field approaches, while having a wider range of validity and requiring much less severe approximations in its treatment of the condensate evolution. 
\par 
The theory presented here contains two additional features that have not yet been investigated in the literature on vortex lattice formation in Bose-Einstein condensates: thermal noise and number conserving scattering. 
While we have not discussed the influence of scattering on vortex dynamics here, 
one may expect that vortices will act as strong scatterers, and indeed that this process may be far more significant for vortex lattices than the weak renormalisation of the growth rate observed for vortex free condensates~\cite{Anglin1999}.
Thermal noise is an essential feature of any dissipative theory, and its effect on vortex lattice formation, both for the mechanical stirring and defect pinning nucleation mechanisms, will be the subject of future work.
\par
AB thanks Matthew Davis, Murray Olsen and Peter Drummond for useful discussions.
This work was supported by \TAD, \ARC, the \VSO, and the \CWGFund.
\section*{References}

\end{document}